\documentclass[showkeys,aps,pre,longbibliography,reprint,superscriptaddress,floatfix,notitlepage]{revtex4-2}

\usepackage[dvipsnames]{xcolor}
\definecolor{linkcolor}{rgb}{0.3,0.3,1.0} 
\usepackage[pdftex,colorlinks=true, linkcolor= linkcolor, citecolor= linkcolor, urlcolor= linkcolor, hyperindex=true,hyperfigures=true]{hyperref} 
\usepackage{amsmath,amssymb}
\usepackage{graphicx}
\usepackage{color}
\usepackage{subfigure}
\usepackage{svg}

\begin{document}

\title{Random Sequential Adsorption with Correlated Defects: A Series Expansion Approach}

\author{G Palacios}
\affiliation{Departamento de Física, Universidade Federal de Pernambuco, 50670-901, Recife, PE, Brazil}
\email{palaciosg226@gmail.com}

\author{A M S Mac\^edo}
\affiliation{Departamento de Física, Universidade Federal de Pernambuco, 50670-901, Recife, PE, Brazil}

\author{Sumanta Kundu}
\affiliation{Scuola Internazionale Superiore di Studi Avanzati (SISSA), Via Bonomea 265, 34136 Trieste, Italy}

\author{M A F Gomes}
\affiliation{Departamento de Física, Universidade Federal de Pernambuco, 50670-901, Recife, PE, Brazil}


\begin{abstract}
The Random Sequential Adsorption (RSA) problem holds crucial theoretical and practical significance, serving as a pivotal framework for understanding and optimizing particle packing in various scientific and technological applications. Here the problem of the one-dimensional RSA of k-mers onto a substrate with correlated defects controlled by uniform and power-law distributions is theoretically investigated: the coverage fraction is obtained as a function of the density of defects and several scaling laws are examined. The results are compared with extensive Monte Carlo simulations and more traditional methods based on master equations. Emphasis is given in elucidating the scaling behavior of the fluctuations of the coverage fraction. The phenomenon of universality breaking and the issues of conventional gaussian fluctuations and the L{\'e}vy type fluctuations from a simple perspective, relying on the Central Limit Theorem, are also addressed.
\end{abstract}

\maketitle


\section{\label{sec:level1}Introduction}

Random Sequential Adsorption (RSA) has garnered significant interest in the study of complex systems, with applications spanning a diverse range of fields from surface physics to molecular biology \cite{talbot2000car, zhang2010molecular,adamczyk2012modeling}. In classical RSA \cite{Feder1980}, particles are added sequentially and irreversibly to a substrate without allowing overlap, resulting in the formation of ordered and highly compact structures. The dynamics of adsorption ceases when the system reaches the jammed state characterized by the maximum coverage of the surface where no more objects can be accommodated. 

The jamming coverage is found to be dependent on the geometry of the substrate, the shape and size of the adsorbed objects, and also, on the underlying dynamics of the adsorption \cite{hinrichsen1986geometry,  albano1993adsorption, budinski2017particle, ciesla2015shapes, kasperek2018random,garcia2015random,ramirez2023random,Krapivsky_2023}.  However, intriguingly, the critical behavior of the system associated with the jamming transition point is observed to be universal. This implies that it is defined by a universal exponent $\nu$, which governs the size scaling of the transition zone width $\sigma$, or standard deviation of jamming coverage. Specifically, $\sigma$ scales with the linear size $L$ of the substrate in a spatial dimension $D$ as $\sigma \sim L^{-1/\nu}$, where $\nu = 2/D$ \cite{pasinetti2019random,ramirez2019inverse}. 
Using heuristic mean-field arguments, as discussed below, we can demonstrate this result. We confine ourselves to the case of $D=1$, that is, one-dimensional systems of size $L$, as is the case here. We define the system's free energy in the usual manner, $F = E - TS$, where $E$ represents the internal energy, $T$ stands for the temperature (a fugacity in this non-equilibrium process), and $S$ denotes the entropy: $S=C\ln\Omega$, with $C$ assuming the role of the Boltzmann constant, and $\Omega$ representing the number of effective microscopic accessible states. Let us consider the following arguments: (i) $E$ in the case of our $k$-mers RSA problem is a self-avoidance energy. It is defined from the two-body approximation: $E \sim d^2L$, where $d$ is the mean-field linear density equal to $M/L$, and $M$ is the associated mass of the system. The $L$ factor in the expression for $E$ stands for integration in the 1D volume. In summary, $E = M^2/L$. (ii) The system's mass, $M$, is defined as $M = \sigma L$, where $\sigma$ is assumed to be proportional to the transition zone width. Then, $E = \sigma^2L$. (iii) If $\Omega \sim L$, the entropic term reads $S = C\ln L$. Utilizing these results and minimizing the free energy with respect to $L$, we finally obtain that $\sigma \sim \sqrt{CT/L}$.

Although the effect of particle properties (such as shape, size and orientation) on the kinetics of adsorption has been extensively studied in the past, the influence of substrate heterogeneities has remained largely unexplored. Interestingly, in numerous real-world systems, the presence of defects on the substrate can profoundly influence the adsorption dynamics and jammed properties. These substrate heterogeneities can significantly alter the adsorption process, leading to non-trivial behaviors that are not captured by traditional models. The only extensively explored case in the literature is when the defects are randomly placed without any spatial correlation \cite{wang1993locally, ben1994irreversible, tarasevich2015impact, centres2015percolation, budinski2016jamming, tarasevich2017influence}. Even in this scenario, the universality associated with the critical exponent $\nu$ still holds true.

Striking examples of the effect that a structured substrate with defects can have on the dynamic and saturation properties of the RSA model were previously studied \cite{Kundu2021, palacios2020random} . These works depart from the tradition of considering only spatially uncorrelated defects. In \cite{palacios2020random}, it is shown that a deterministic organization of the defects leads to an unexpected efficiency effect on the coverage fraction. On the other hand, in \cite{Kundu2021}, it is demonstrated that considering a spatial correlation following a power-law leads to a breakdown of the universality of the exponent $\nu$.

In this study, we explore the problem of RSA while considering spatial correlation among defects. To address this, we propose introducing correlation by assuming that defect placement follows a one-dimensional and unidirectional random walk. We consider two cases: one where the step size follows a uniform distribution and another where the step size follows a power-law distribution. Finally, we offer an explanation for the universality breaking problem discussed earlier by delving into the scaling behavior of coverage fraction fluctuations. Our findings suggest that our system falls within a class of complex systems where a transition between Gaussian and L{\'e}vy statistics emerge \cite{mantegna1994stochastic, perc2007transition, palacios2023replica}.

\section{Theoretical Framework}
The one-dimensional RSA problem involving random spatial uncorrelated defects was analytically addressed in \cite{ben1994irreversible}, while the two-dimensional counterpart was tackled through Monte Carlo simulations \cite{centres2015percolation, budinski2016jamming, lebovka2015jamming, palacios2020random,Kundu2021,Kundu2022}. The analytical solution for the 1D scenario leverages the notion that random defects within a substrate can be interpreted as a Markov chain, thus permitting the utilization of the same rate equations as those used for the original RSA problem. This solution, governing the deposition of defects, forms the foundation for establishing an initial condition for the subsequent determination of the appropriate solution for $k$-mers adsorption \cite{ben1994irreversible}. The ultimate expression for the coverage fraction $\theta$ in relation to the initial density of defects $\theta_0$ is derived through the equation:

\begin{equation}
  \theta(k,\theta_0) = k(1-\theta_0)^k \int_{0}^{1} e^{-2\sum_{i=1}^{k-1} \frac{(1-x^i)(1-\theta_0)^i}{i}}dx.
  \label{eq:defrandom}
\end{equation}

The challenge of describing the RSA problem with correlated defects using a methodology grounded in master equations stems from the intrinsic characteristics of the stochastic processes employed to place defects within the substrate while accounting for correlations. 
The conventional methods based on master equations assume that both the adsorption of defects and particles are encompassed within the same theoretical framework, treating adsorption as a Poisson process in which all substrate sites possess equal probability for adsorption.
Consequently, any solution aiming to address the adsorption issue in the presence of correlated defects must possess the capability to distinctly separate the process of defect adsorption from particle adsorption. In the forthcoming sections, we outline the fundamental principles of our model tailored to address random sequential adsorption in the context of spatial correlated defects.

In one dimension, the presence of defects can be conceptualized as the fragmentation of the substrate. In this scenario, the gap between two adjacent defects forms an empty space or ``fragment" where the posterior adsorption of $k$-mers takes place independently from other fragments. To holistically address the problem, two key aspects must be addressed: (i) the determination of the average number of $k$-mers that can be deposited within a finite fragment devoid of defects, and (ii) the distribution of fragment sizes. The recursive approach introduced subsequently represents a promising methodology to address the first aspect. 

\begin{figure} [h]
\includegraphics [width=.8\columnwidth]{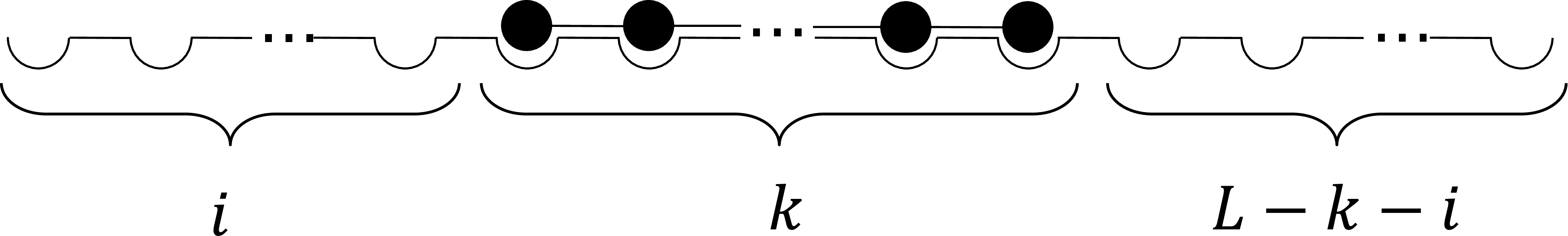}
\caption{Adsorption of first $k$-mer.}
\label{fig:recursivemethods}
\end{figure}

To derive the recurrence equation, suppose  that we have a finite one-dimensional substrate, with $L \geq k$ sites, where $k$-mers can be absorbed. The first $k$-mer can be randomly placed in the first $L-k$ sites, after which its position is fixed, and any overlap of $k$-mers is prohibited. Therefore, when the first $k$-mer is adsorbed, two independent gaps appear with $i$ and $L-k-i$ sites respectively, as seen in Fig \ref{fig:recursivemethods}.

Let $a_L^k$ be the expected number of $k$-mers, in mean, that can eventually occupy a chain of $L$ initially empty sites, after the adsorption of the first $k$-mer occupying positions $i+1,\ldots,i+k+1$. The expectation will be the sum of the expectations of the three intervals now created, namely $\langle a_i^k + a_k^k + a_{L-i-k}^k \rangle_i$. Taking the average over all possible values of $i$ (with uniform distribution), and considering that the first and last intervals are symmetric, we finally have:


\begin{equation}
    a_L^k = 1 + \frac{2}{L-k+1}\sum_{i=1}^{L-k} a_i^k,
\end{equation}
which can be written in recursive form as \cite{palacios2020random}:

\begin{equation}
    (L+1)a_{L+k}^k-La_{L+k-1}^k-2a_{L}^k=1.
    \label{eq:recurrence}
\end{equation}

To solve this recurrence equation we can apply the Z-transform on both side of (\ref{eq:recurrence}), then, we obtain a ordinary differential equation for $X(z)=\mathcal{Z}[a_L^k]$: 

\begin{eqnarray}
    \frac{dX}{dz} + \frac{(k-1)z^k- z^{k-1}(k-1)+2}{z^k(z-1)} X  \nonumber \\
    = - \frac{1}{z^{k-1} (z-1)^2}, 
    \label{eq:recurrenceODE}
\end{eqnarray}
whose general solution, after some algebric     manipulation, is given by,


 
\begin{eqnarray}
    X(z) = \frac{\exp{\left[-2 \sum_{i=1}^{k-1} \frac{z^{i-2}}{i}\right]}}{z^{k-3}(z-1)^2} \times \nonumber \\
    \left[C - \left[\int{\exp{\left(2\sum_{i=1}^{k-1} \frac{x^{2-i}}{i}\right)}dx}\right]_{x=\frac{1}{z}}\right].
    \label{eq:ODEsolution}
\end{eqnarray}

For example, for dimers ($k=2$), the integral in (\ref{eq:ODEsolution}) can be explicitly calculated leading to,

\begin{eqnarray}
    X(z) = \frac{1}{2} \frac{z}{(z-1)^2} + C\frac{z \exp{\left(-\frac{2}{z}\right)}}{(z-1)^2}.
    \label{eq:ODEsolutionk2}
\end{eqnarray}

Applying the inverse Z-transform and using the initial condition $a_{L=2}^{k=2}=1$ we get

\begin{eqnarray}
    a_L^{k=2} = \mathcal{Z}^{-1}[X(z)] = \frac{L}{2} - \frac{(-1)^{L+1} 2^{L-1}}{2(L-1)!} \times \nonumber \\
    \times\Theta(L-1){}_2F_0(2,1-L;;\frac{1}{2}),
    \label{eq:ODEsolutionk2_1}
\end{eqnarray}
where ${}_2F_0(\cdot)$ is the generalized hypergeometric function and $\Theta(\cdot)$ is the Heaviside step function, with the convention $\Theta(0)=1/2.$

Now, the key point lies in determining the distribution of fragment sizes arising from defects present within the substrate. This task is specific for each spatial distribution of defects, rendering the effectiveness of the proposed method contingent on the ability to derive this distribution function from the defect arrangement.

Therefore, the average number of $k$-mers that can be deposited on a substrate of $L$ sites initially occupied by a density $\theta_0$ of defects, $M(\theta_0,L,k)$, will be: the sum over all possible values of the fragment size $\eta$ of the product between the the average number of $k$-mers that can eventually be deposited on a fragment with $\eta$ sites ($a_{\eta}^k$), times the number of fragments of type $\eta$, i.e., $n_{\eta}$, so,

\begin{equation}
    M(\theta_0,L,k) =\sum_{\eta} n_{\eta}a_{\eta}^k,
\end{equation}
or, in terms of the probability distributions $P(\eta)$,


\begin{equation}
    M(\theta_0,L,k) = N \sum_{\eta} P(\eta)a_{\eta}^k,
    \label{eq:method}
\end{equation}
where \(N=\theta_0 L\) represents the total number of fragments.

Indeed, it's apparent that eq. (\ref{eq:method}) presents a broader formulation compared to (\ref{eq:defrandom}), which is exclusively applicable to a random and uncorrelated defect distribution. Ultimately, the coverage fraction, defined as $kM/L$, can be expressed as:

\begin{equation}
    \theta(\theta_0,L,k) = k\theta_0 \sum_{\eta} P(\eta)a_{\eta}^k.
    \label{eq:theta}
\end{equation}

Note that we only have a closed form of (\ref{eq:theta}) for dimers. However, the numerical solution of (\ref{eq:theta}) for $k>2$, using recurrence (\ref{eq:recurrence}) is much more efficient than the Monte Carlo simulation, that would be the traditional way to solve the problem of RSA with correlated defects.

\section{Results}
In this section we test the eq. (\ref{eq:theta}) for different types of distribution of defects, including Random (uncorrelated) distribution (to compare with (\ref{eq:defrandom})), Power law and Uniform distribution of the size of fragments. 

\subsection{Random (uncorrelated) defects}
In this initial example of our proposed method, our objective is to demonstrate the applicability of model (\ref{eq:theta}) to a well-known scenario. Specifically, for the case of a random defect distribution, computing $P(\eta)$ is straightforward. Placing defects randomly within the substrate can be linked to a Bernoulli process, where $p$ represents the probability of success (defect presence) and $q=1 - p$ is the probability of failure (site being unoccupied). It is not difficult to conclude that $p=\theta_0$. In this context,

\begin{equation}
    P(\eta) = \theta_0(1-\theta_0)...(1-\theta_0)\theta_0 = \theta_0^2(1-\theta_0)^\eta.
\end{equation}

Equation (\ref{eq:theta}) imposes that $P(\eta)$ is marginally normalized in $\eta$ with $\frac{1}{\theta_0}$ being the normalization factor. When $L$ is sufficiently large we can neglect the boundary effect, and the expression for the coverage fraction is written as:

\begin{equation}
    \theta(\theta_0,L,k) = k\theta_0^2 \sum_{\eta=0}^{\infty} (1-\theta_0)^\eta a_{\eta}^k.
    \label{eq:thetarandom}
\end{equation}

\begin{figure} [h]
\includegraphics [width=.8\columnwidth]{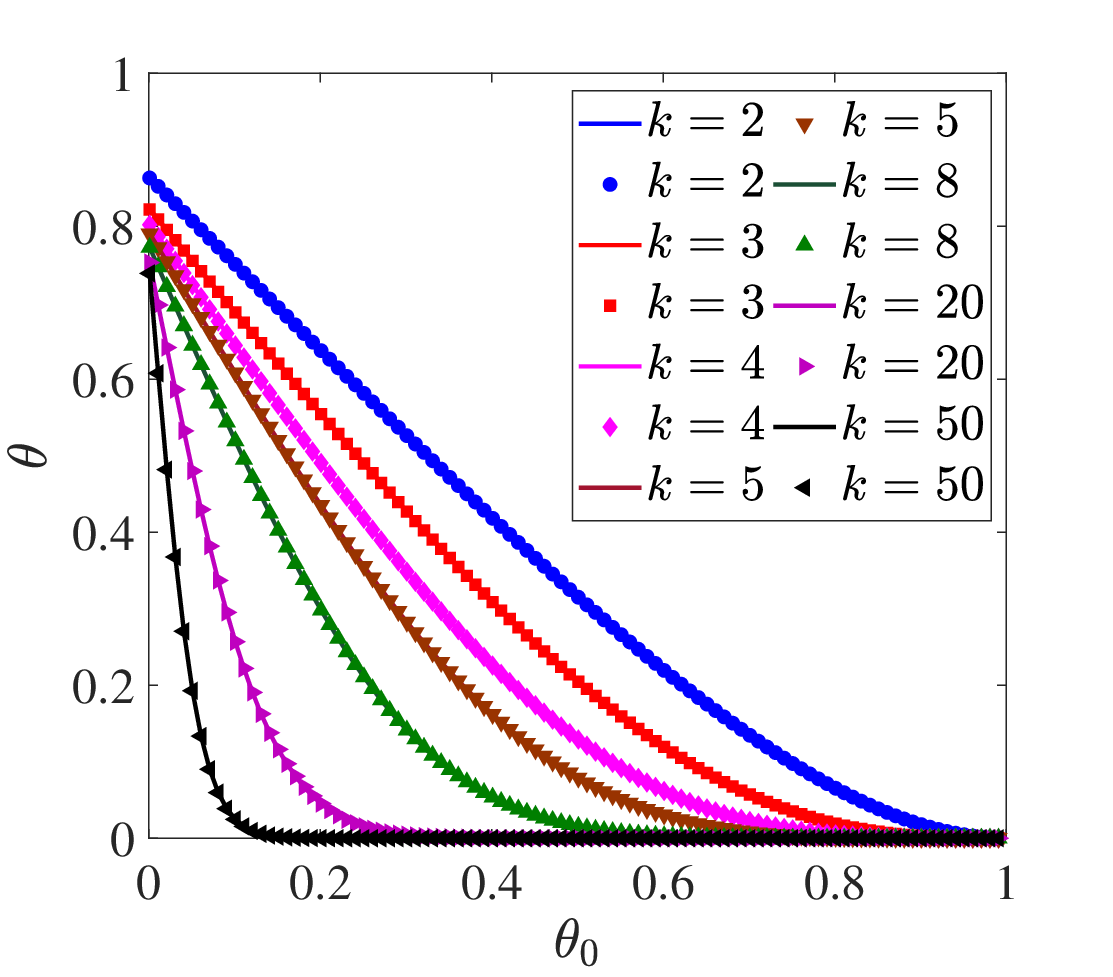}
\caption{\label{fig:wide} Comparison of the results of the numerical solution (using the trapezoidal method) of (\ref{eq:defrandom}) (symbols) and (\ref{eq:thetarandom}) (solid lines) for different values of $k$.}
\label{fig:random}
\end{figure}

To prove the validity of (\ref{eq:thetarandom}), in Fig. \ref{fig:random} we show a comparison between the graphs of equations (\ref{eq:defrandom}) and (\ref{eq:thetarandom}), the first being the standard model for the treatment of the RSA problem with random defects. For the case of $k=2$, the solid blue line in Fig. \ref{fig:random}, we have used the closed form solution for $a_{\eta}^{k=2}$ obtained in (\ref{eq:ODEsolutionk2_1}).

The correspondence between the results depicted in Fig. \ref{fig:random} should not come as a surprise. It is evident that (\ref{eq:thetarandom}) constitutes the Z-transform of the sequence $a_{\eta}^k$, multiplied by the factor $k\theta_0^2$, where the parameter $z=\frac{1}{1-\theta_0}$. For instance, we can verify that under these conditions and for especific case of dimers ($k=2$) the equation (\ref{eq:ODEsolutionk2}) reduces to:
\begin{equation}
X\left(z=\frac{1}{1-\theta_0}\right)=(1-\theta_0)\left[1-\exp{(-2(1-\theta_0))}\right],
\end{equation}
which is consistent with \cite{ben1994irreversible}. The reader can easily verify that for any $k$, (\ref{eq:thetarandom}) reduces to (\ref{eq:defrandom}) under these considerations.

Although our model (\ref{eq:thetarandom}) does not add considerable advantage or predict new phenomena, in the specific case of the random distribution of defects, the mathematical structure is certainly much simpler and more intuitive than the dynamic approach in terms of master equations. Consider that in model (\ref{eq:thetarandom}) finite size effects are present, this could be considered as a disadvantage with respect to the analysis from the kinetic point of view of adsorption, equation (\ref{eq:defrandom}). However, we must remember that the latter only describes infinitely large systems, whereas real systems are finite.

\subsection{Spatially correlated defects}

Suppose that the defects are placed on the substrate in a manner where the positions $x_i$ are determined by:

\begin{equation}
    x_{i}=x_{i-1}+r,  x_i<L.
    \label{eq:proces}
\end{equation}

Here $r$ is a random variable with a given probability distribution $Q(r)$ over the closed interval $\left[r_{min},r_{max}\right]$. We will consider that $r\geq1$ and $x_0=0$. This represents the problem of a one-dimensional random walker (unidirectional) with a step size distribution $Q(r)$. We will also consider that the discrete count $i$, ends for a given $i_f$ such that $x_{i_f} \leqslant L$. The defective substrate resulting from the application of the stochastic process (\ref{eq:proces}) is where the deposition of the $k$-mers will take place. For the deposition of $k$-mers, we will consider hard wall boundary conditions.


The defect density on the substrate is a random variable, and its distribution function has a closed analytical form that is very challenging to derive. Nevertheless, an estimate for the mean value of the defect density can be obtained for a sufficiently large $L$ as:

\begin{equation}
    \langle \theta_0 \rangle = \frac{1}{\langle r \rangle}.
    \label{eq:estimatheta0}
\end{equation}
Consequently, we rewrite Eq. (\ref{eq:theta}) as:

\begin{equation}
    \theta(\theta_0,L,k) = k\langle\theta_0 \rangle \sum_{\eta} P(\eta)a_{\eta}.
    \label{eq:theta1}
\end{equation}

\subsubsection{Uniform distribution}

Assume that \(Q(r)\) is a uniform distribution in the domain $r \in [2,r_{max}]$ that implies $\eta \in [1,r_{max}-1]$, so: 

\begin{equation}
    Q(r) = \frac{1}{r_{max}},
\end{equation}
leading to

\begin{equation}
    \langle r \rangle = \frac{1}{2}(r_{max}+1).
\end{equation}
Finally,

\begin{equation}
    \theta(\theta_0,L,k) = \frac{2k}{r_{max}(r_{max}+1)} \sum_{\eta=1}^{r_{max}-1} a_{\eta}^k.
    \label{eq:thetauniform}
\end{equation}

\begin{figure} [!h]
\includegraphics [width=.9\columnwidth]{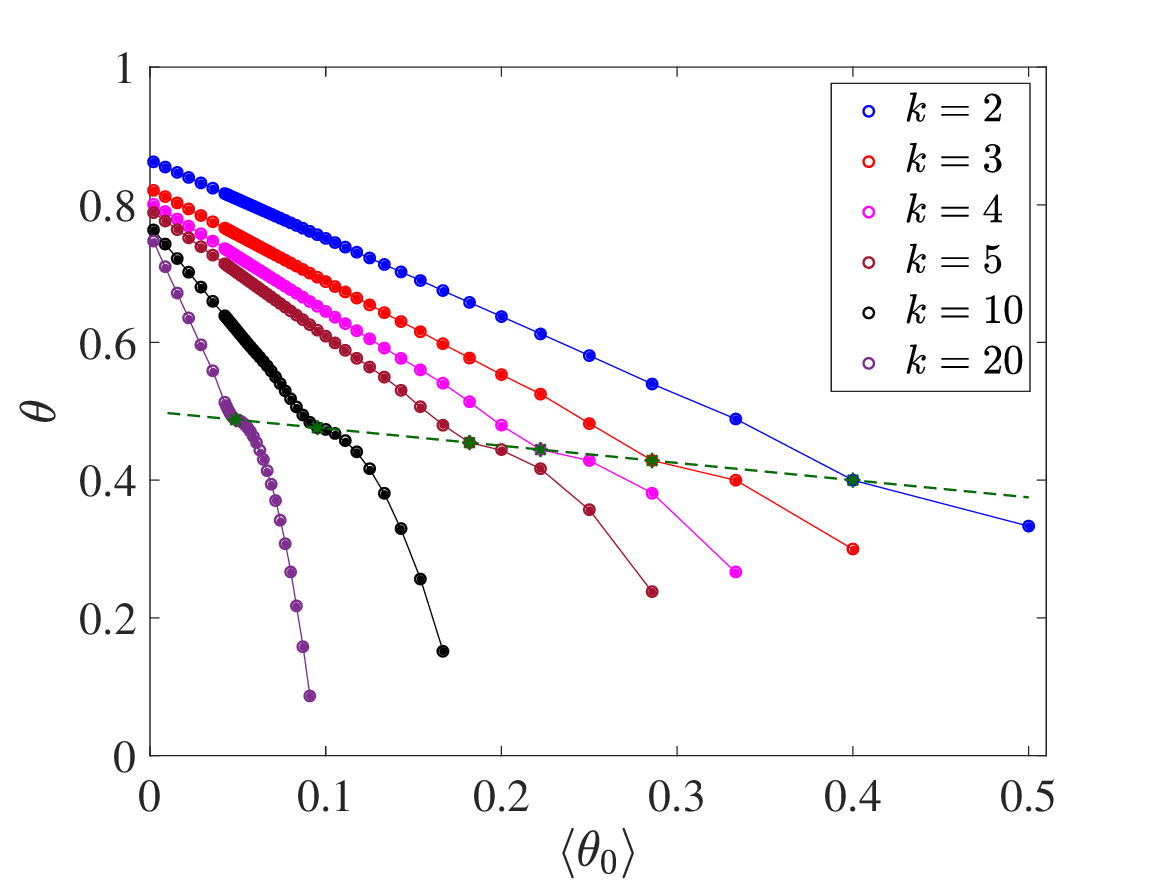} \\
\caption{\label{fig:wide}Results for the coverage fraction as a function of density of defects for the case of uniformly distributed defects. We compare the results obtained from eq. (\ref{eq:thetauniform})  (empty circles) and Monte Carlo simulation (filled circles) for several values of $k$. The dashed straight line connects the points where
$\theta(k,\langle\theta_0\rangle)$ changes concavity.}
\label{fig:uniform}
\end{figure}

In Fig. \ref{fig:uniform} we show the dependence between the coverage fraction and the mean density of defects obtained using the eq. (\ref{eq:thetauniform}) and Monte Carlo simulation.

It is worth noting in Fig. \ref{fig:uniform} that, unlike the random case (Fig. \ref{fig:random}), the coverage fraction does not only decrease while maintaining a concave behavior for all values of $k$ and $\theta_0$. At specific points, precisely where $\langle r \rangle = 2k$ implying $\langle\theta_0\rangle(2k)=\langle\theta_0\rangle'$, there is a change in the slope of the curve $\theta(\langle\theta_0\rangle)$, transitioning to a convex behavior. A straight line passing through these points can be expressed by the simple equation: $\theta(\langle\theta_0\rangle') = 1/2 - 1/4 \langle\theta_0\rangle'$ (dashed green line in Fig. \ref{fig:uniform}).



\subsubsection{Power law distribution}
If we assume a discrete power law distribution, then:

\begin{equation}
    Q(r) = \frac{r^{-\alpha}}{\sum_{r=2}^{r_{max}}r^{-\alpha}}.
    \label{eq:powerlaw}
\end{equation}

The distribution (\ref{eq:powerlaw}) is known as generalized Zipf's law \cite{mandelbrot1982fractal}. If the separation distance between two defects in a lattice follows a Zipf's law, it could indicate a long-range interaction between the defects. Zipf's law implies that short distances, i.e. small $r$’s occur frequently, while large values of $r$ are rare. In the context of the interaction between defects in a lattice, this suggests that some defects may be strongly correlated at long distances, which could be characteristic of long-range interactions or non-local correlations among defects in the lattice.

We will analyze the case in which $r$ is in the interval $[2,r_{max}=L]$. Note that in this case, $\eta \in [1, L-1]$, implying the original format of Zipf's law where the frequency $f(\eta) \sim \eta^{-1}$, with $\eta$ being the rank of the variable and $\eta=1$ for the most frequent outcome \cite{mandelbrot1982fractal}. So, the density of defects reads as: 

\begin{equation}
    \langle\theta_0\rangle = \frac{1}{\langle r \rangle} = \frac{\sum_{r=2}^{L}r^{-\alpha}}{\sum_{r=2}^{L}r^{-\alpha+1}}.
    \label{eq:theta0vsalpha}
\end{equation}

\begin{figure} [!h]
\includegraphics [width=.9\columnwidth]{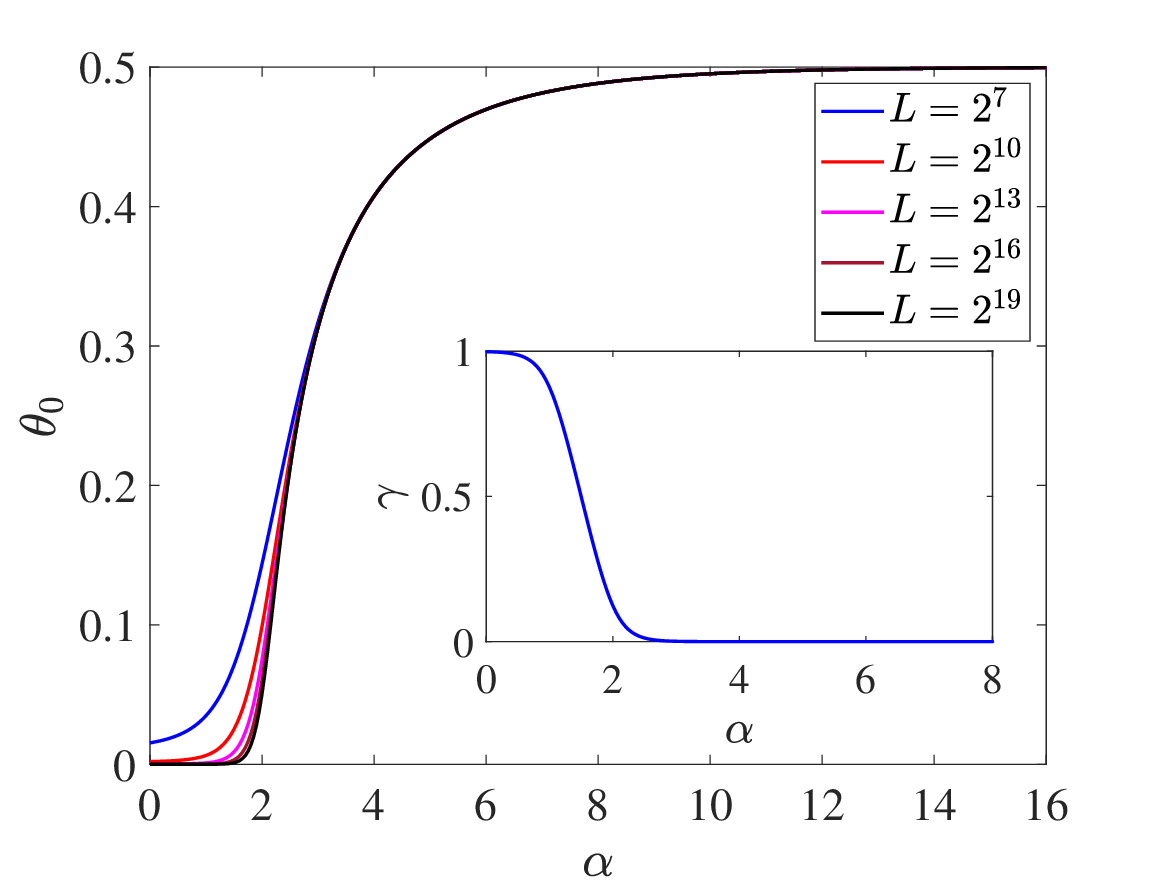}
\caption{Density of defects as a function of the exponent $\alpha$ for various system sizes $L$ (equation (\ref{eq:theta0vsalpha})). Inset: Plot of the exponent of the scaling behavior $\theta_0(\alpha, L) \sim L^{-\gamma}$ for all values of $\alpha$ examined. }
\label{fig:theta0vsalpha}
\end{figure}

In Fig. \ref{fig:theta0vsalpha}, we present the plot of Eq. (\ref{eq:theta0vsalpha}) for different values of $L$. 

An interesting observation from Fig. \ref{fig:theta0vsalpha} is that when the defects are arranged in this manner, a state change occurs. It is noteworthy that for $ \alpha < 2 $, the defect density is practically zero. However, for \( \alpha > 2 \), there is a sudden jump in the defect density. This behavior becomes more pronounced as $L$ increases. To analyze the scaling behavior, for different values of $\alpha$, we fit the model $\theta_0(\alpha, L) \sim L^{-\gamma}$, and the results for the dependence of $\gamma$ vs $\alpha$ are shown in the inset of Fig. \ref{fig:theta0vsalpha}.

\begin{figure} [h!]
\includegraphics [width=.9\columnwidth]{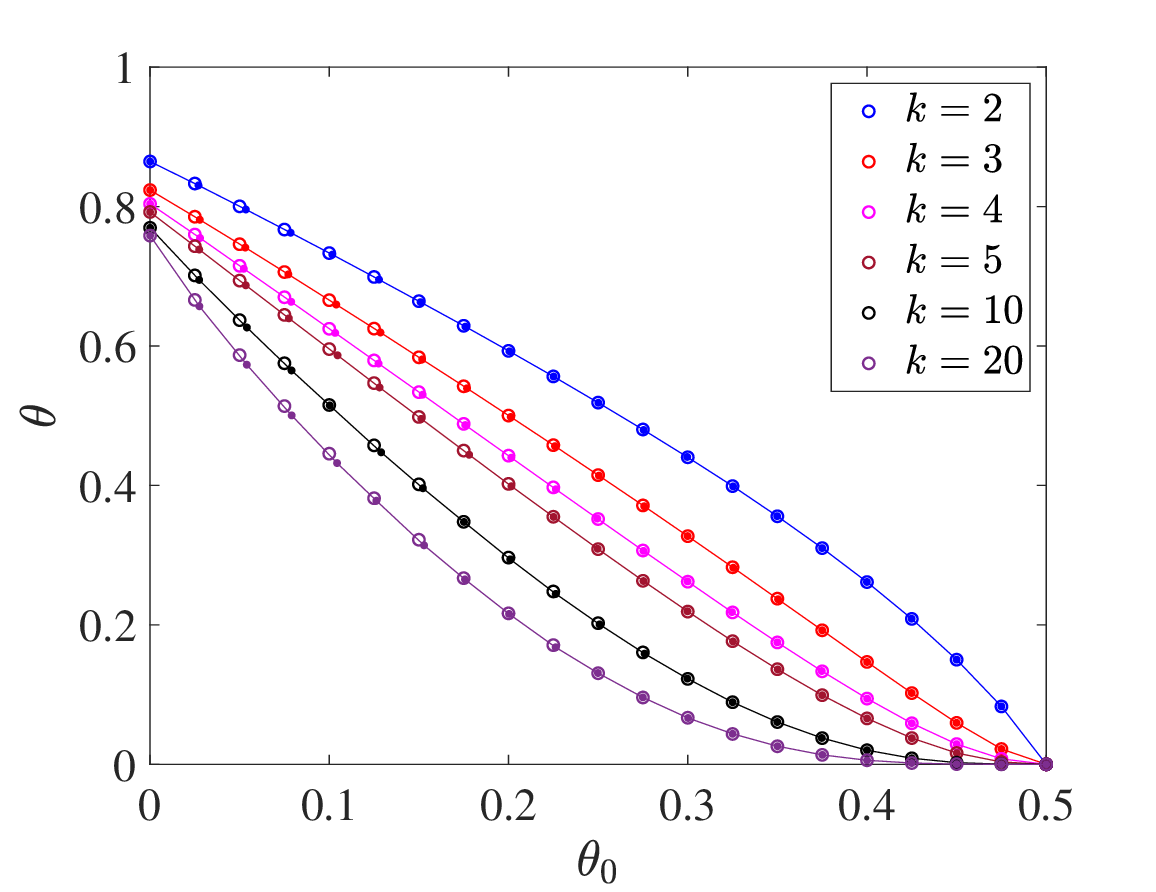}
\caption{\label{fig:wide} Results for the coverage fraction as function of density of defects for the case of a power law distribution. We compare the results obtained with our model  (empty circles) and with Monte Carlo simulation (filled circles) for several values of $k$. We use a lattice size $L = 2^{16}$.}
\label{fig:powlaw}
\end{figure}

Because there is no close functional relationship between exponent $\alpha$ and the density of defects, the coverage fraction curves (Fig. \ref{fig:powlaw}) were obtained by varying $\alpha$. Finally the coverage fraction is:
\begin{equation}
    \theta(\theta_0,L,k) = \frac{k}{\sum_{r=2}^{L}r^{-\alpha+1}} \sum_{r=1}^{L} r^{-\alpha} a_r^k.
\end{equation}

Note in Fig. \ref{fig:powlaw} that the functional relation $\theta(\langle \theta_0 \rangle)$ presents a transition from convex for $k=2$ to concave for $k \geqslant 4$, passing through an almost linear dependence in the case of $k=3$. It is interesting to mention that in a previous work \cite{palacios2020random} we had already observed a similar behavior, in which the adsorption of dimers seems to have a completely different behavior compared to the rest of the $k$-mers ($k\geq3$) when varying the defect density. In the case of \cite{palacios2020random}, we found that the curve $\theta(\langle \theta_0 \rangle)$ for dimers does not fit well together with the other curves ($k \geq 3$) when performing a finite-size analysis with the curve collapse method. In the current study, once again we have encountered a case where the adsorption of dimers exhibits a distinctive behavior.

\subsection{Fluctuations}
At this point, the reader can see that all fluctuations associated with the coverage fraction are linked to the stochastic behavior of defect placement on the substrate. This means that we will assume, by hypothesis, that the fluctuations of the coverage fraction exhibit the same scaling behavior of the fluctuations of $\theta_0$. 

To begin our approach, we want to calculate the probability distribution $P(N)$ of the maximum number $N$ of steps in the walk such that in step $N+1$ the walker exits the lattice. Mathematically, we write:
\begin{equation}
   P(N) = P(S_N \leq L \,|\, S_{N+1} > L). 
\end{equation}
This conditional probability can be written as:

\begin{equation}
P(N) = P(S_N=L) + \sum_{i=1}^{r_{max}-1}\sum_{j=i+1}^{r_{max}} P(S_N=L-i)P(r=j).
\label{eq:PN}
\end{equation}

In other words, we aim to calculate the probability that, until the $N$-th draw, the total sum $S_N$ is equal to or less than $L-i$ ($i \in [0,r_{max}-1]$), and that in the $(N+1)$-th draw the sum $S_{N+1}$ exceeds $L$. To achieve this, we must consider, for each $L-i$, all possible cases in which the next draw results in a sum exceeding $L$. Since these are independent probabilities, we simply take the product of each of them. Scale arguments can be used to simplify (\ref{eq:PN}), reducing it to:

\begin{equation}
P(N) \approx P(S_N=L).
\label{eq:PNaprox}
\end{equation}

On the other hand, the moment generating function (mgf) of the discrete random variable $r$ is defined as the expected value of the function $e^{tr}$ \cite{feller1991introduction}. If we sum the independent random variables $r_1, r_2, ..., r_N$, the mgf of the sum $S_N$ is :

\begin{equation}
    M_{S_N}(t) = \left[\sum_{r=1}^{r_{max}}e^{tr}Q(r)\right]^N = \sum_{s=N}^{Nr_{max}} e^{ts}  P(S_N=s).
    \label{eq:MSN}
\end{equation}

For the case where $Q(r)$ is uniform, we can identify in equation (\ref{eq:MSN}) that:

\begin{equation}
    P(S_N=L) = \frac{1}{r_{max}^N} \sum_{k_1+2k_2+...+r_{max}k_{r_{max}}=L} \binom{N}{k_1,...,k_{r_{max}}}.
\end{equation}

In other words, the probability of $S_N=L$ will be the sum of all coefficients of the polynomial (\ref{eq:MSN}) whose variable is $\left(e^t\right)^L$, represented by the multinomial expansion coefficient $\binom{N}{k_1,...,k_{r_{\text{max}}}}$ satisfying the condition $k_1+2k_2+\ldots+r_{\text{max}}k_{r_{\text{max}}}=L$. By employing some numerical algorithm, $P(S_N = L)$ can be determined exactly, although as $N$ increases, the calculation becomes increasingly cumbersome. However, Gaussian behavior emerges from the Central Limit Theorem (CLT), asserting that the sum of $N$ random variables $\{r\}$ that are statistically independent, identically distributed, with mean $\mu_r$ and with finite variance $\sigma_r$ converges (as $N \to \infty$) to a normal (Gaussian) distribution with mean $\mu_{S_N}=N\mu_r$ and variance $\sigma_{S_N}^2=N\sigma_r^2$. Consequently, the distribution $P(S_N)$ can be approximated as:

\begin{equation}
    P(S_N) = \frac{1}{\sqrt{2\pi N\sigma_r^2}} \exp{\left[-\frac{1}{2}  \frac{\left( S_N-N\mu_r\right)^2}{N\sigma_r^2}\right]}.
    \label{eq:PSN}
\end{equation}
Assuming the approximation (\ref{eq:PNaprox}) and rearranging some terms in (\ref{eq:PSN}) we finally find that the probability distribution of the density of defects $\theta_0 = N/L$ is:


\begin{equation}
    P(\theta_0) = \frac{A}{\sqrt{2\pi L \theta_0\sigma_r^2}} \exp{\left[-\frac{1}{2}  \frac{\left(\theta_0 - 1/\mu_r\right)^2}{\frac{1}{L}\theta_0\sigma_r^2/\mu_r^2}\right]},
    \label{eq:PStheta0}
\end{equation}
where the normalization constant $A$ can be  calculated, bearing in mind that $\theta_0$ belongs to the interval $[0,1]$:

\begin{equation}
    A = \frac{2L\mu}{2-e^{\frac{2L\mu}{\sigma^2}}\text{erfc}\left(\frac{\sqrt{L}(\mu+1)}{\sqrt{2}\sigma}\right)-\text{erfc}\left(\frac{\sqrt{L}(\mu-1)}{\sqrt{2}\sigma}\right)}.
    \label{eq:Normalization}
\end{equation}


Using the Mathematica software, it can be demonstrated that, as the limit of $L \to \infty$, the mean of distribution (\ref{eq:PStheta0}) converges to $1/\mu_r$, consistent with the assumption made in estimation (\ref{eq:estimatheta0}). On the other hand, through an asymptotic expansion of the variance of (\ref{eq:PStheta0}), it is established that it scales inversely with $L$, thereby indicating that the standard deviation of the density of defects, and consequently the standard deviation of the coverage fraction of $k$-mers, exhibit a scaling behavior of $L^{-1/2}$. It would not be unwarranted to conjecture that this universal behavior persists in all instances where the step size distribution $Q(r)$ remains within the validity bounds of the CLT.

\begin{figure} [!h]
\includegraphics [width=.49\columnwidth]{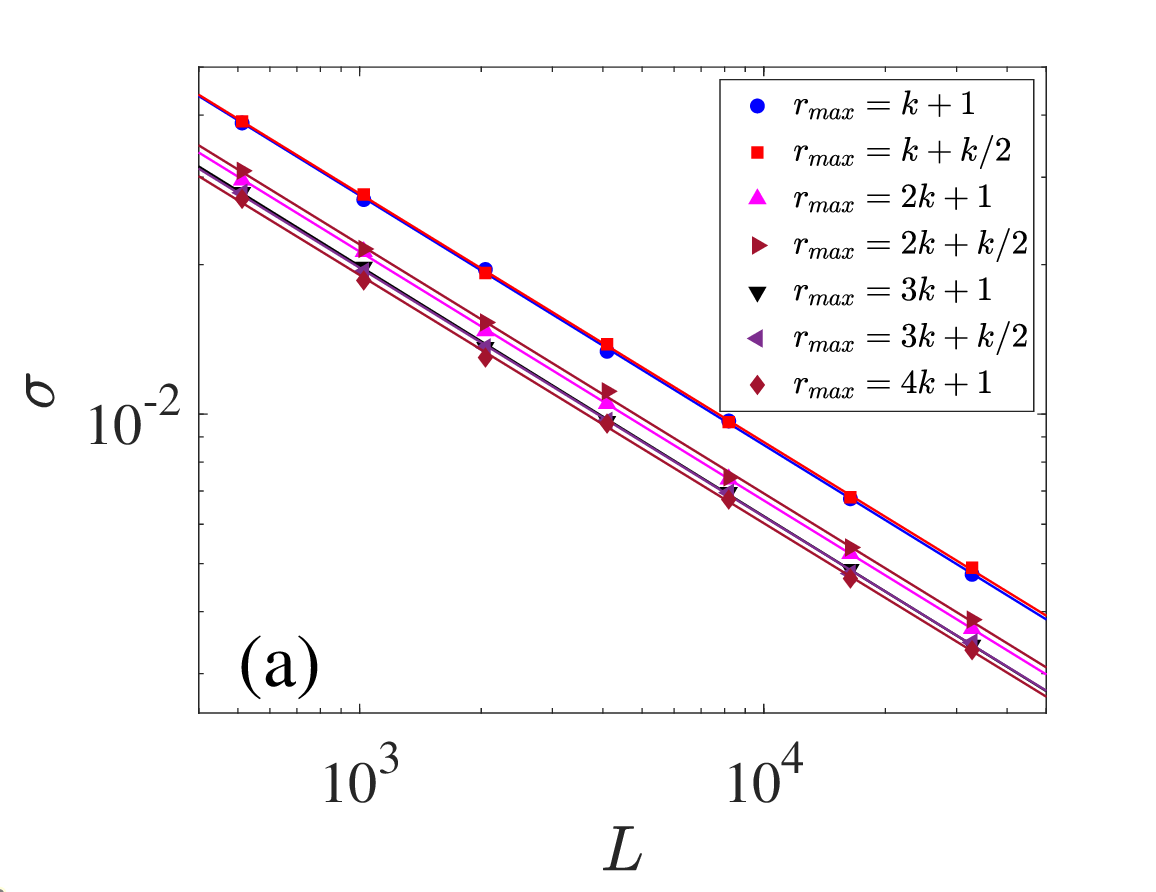}
\includegraphics [width=.49\columnwidth]{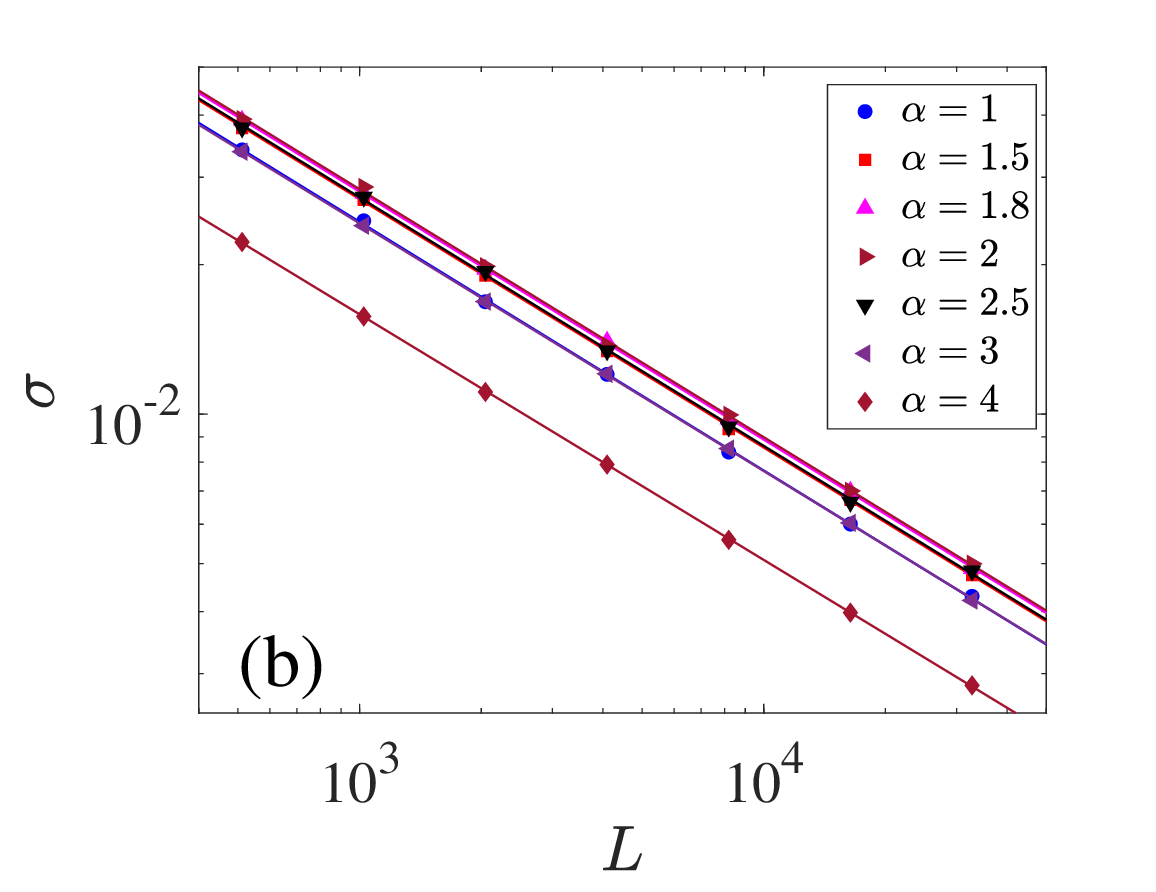}
\caption{Log-log plot illustrating the relationship between the standard deviation of the coverage fraction $\sigma_{\theta}$ and the system size $L$ using Monte Carlo simulations, across various values of: $r_{max}$ for a uniform distribution of defects (a) and $\alpha$ for the power-law distribution of defects (b), with a fixed value of $k=5$. In the case (b) we consider $r_{max} = 10$. The symbols depict simulation data, while the solid lines represent fits to a power-law model of the form $\sigma = AL^{-1/\nu}$. The exponent $\nu$ have a mean value of $1.9983$ and standard deviation of $0.0084$ if we consider all fits.}
\label{fig:sigmarmaxpeque}
\end{figure}

In Fig. \ref{fig:sigmarmaxpeque}, we present the results of numerical simulations, displaying a log-log plot illustrating the scaling behavior of the standard deviation of the coverage fraction $\sigma_{\theta}$ with respect to the system size $L$.

Note that in both cases, we have considered that $r_{max} \ll L$, then, the number of steps needed to stop the stochastic process (\ref{eq:proces}) is considerably larger, causing the system to inevitably fall into the attractor dominated by Gaussian fluctuations. Consequently, as demonstrated in (\ref{eq:PStheta0}), this implies $\nu = 2$.

At this point we remark that, as the statistics of the sum of $N$ independent random variables $r$ with power-law distribution is described \cite{mantegna1994stochastic} by the $\alpha$-stable Lévy distribution whose probability distribution generally does not have a closed form. Generally, the law of large numbers and the central limit theorem may not be directly applicable in this case due to the divergence of the variances. However, by ensuring a cutoff (finite value of $r_{max}$), a finite variance assumption is guaranteed, and convergence to the Gaussian limit can be demonstrated, even though this convergence may be relatively slow \cite{mantegna1994stochastic}. In practical terms, there exists a crossover between Gaussian and anomalous behavior (Lévy regime), which can be characterized by the parameter $N_+$, representing the number of steps required to transition from the Lévy regime to the Gaussian regime. In \cite{mantegna1994stochastic}, it is shown that the scaling relationship is given by $N_+ \sim r_{max}^{\alpha}$.

Consider first the scenario where $r_{max}$ is comparable to the system size. The interpretation of the uniform case is straightforward. In this situation, the walker is permitted to take steps with a length comparable to the lattice size, and as a result, it would reach the boundary with relatively few steps. Consequently, there would be few or almost no defects in the lattice, and thus, we approach the case of a defect-free substrate where fluctuations scale with the universal behavior $(\nu=2)$ \cite{pasinetti2019random}.

Let us now examine the scenario where $Q(r)$ is given by the Zipf's law (\ref{eq:powerlaw}). In this situation, on one hand, small-step lengths remain more probable, allowing the walker to take many steps before reaching the boundary $x=L$. On the other hand, the assumptions of the CLT are not satisfied because as $L \to \infty$, the variance of (\ref{eq:powerlaw}) also tends to $\infty$, leading to an expectation of anomalous behavior in fluctuations.

\begin{figure} [h!]
\includegraphics [width=.49\columnwidth]{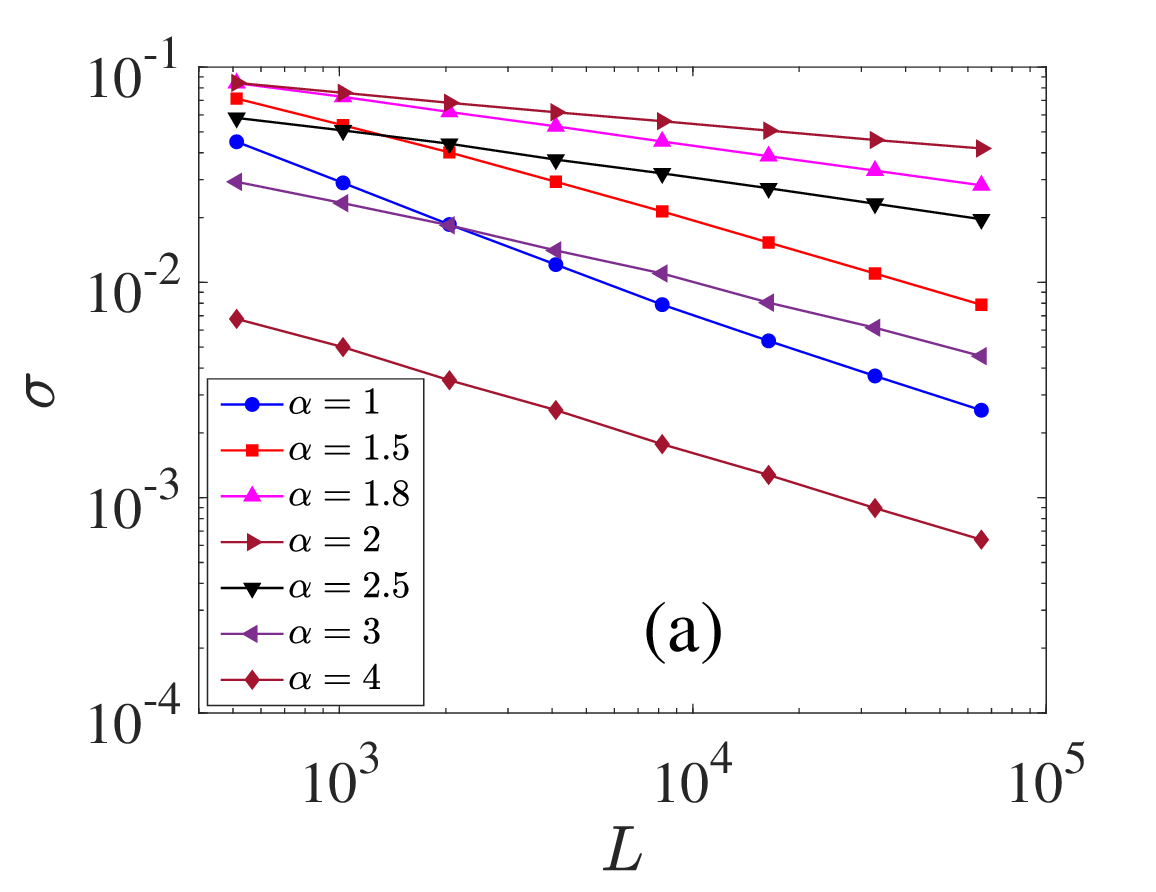}
\includegraphics [width=.49\columnwidth]{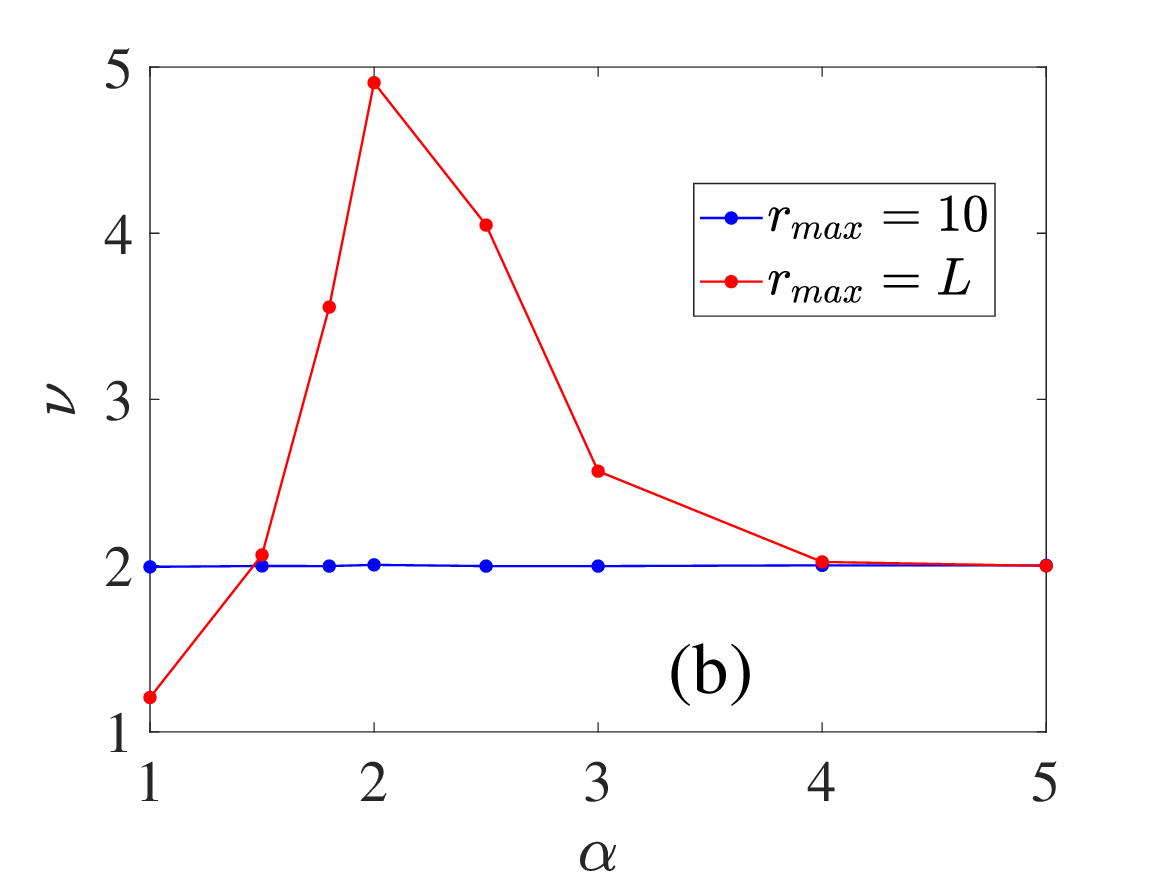}
\caption{\label{fig:wide} (a) Log-log plot of the standard deviation of the coverage fraction $\sigma(L)$ with respect to the system size $L$ for different values of $\alpha$ and for a fixed vale of $k=10$. (b) Scale exponent $\nu$ of the fluctuations (standard deviation) as a function of the exponent $\alpha$, considering cases where $r_{max} = 10 \ll L $ and $r_{max} = L$.}
\label{fig:powlawf}
\end{figure}

In Fig. \ref{fig:powlawf} (a), we present a log-log plot illustrating the scaling behavior of the standard deviation of the coverage fraction ($\sigma$) with respect to the system size ($L$). In Fig. \ref{fig:powlawf} (b), the influence of the maximum step length $r_{max}$ on the scale exponent of fluctuations $\nu$ is presented. It can be observed that when $r_{max} = L$, the number of steps required to halt the stochastic process (\ref{eq:proces}) is very small, specifically $N<N_+$, placing the system in the anomalous region where Levy-type fluctuations dominate, leading to the breakdown of universality. Note in Fig. \ref{fig:powlawf} that for $\alpha = 1.5$ and $\alpha > 4$, $\nu=2$, and $\nu$ attains its maximum value when $\alpha = 2$. \\

\section{Conclusions}
The significance of advancing the understanding of the RSA model with correlated defects cannot be underestimated. In addition to its general intrinsic theoretical interest for statistical physics and for the vast universe of Lévy statistics, in particular, a correct explanation of this model enables improvements in applications across various branches of technology. This is achieved by designing a well-structured defect landscape, allowing the enhancement of specific properties in a material for customized uses.

In this study, we explored the RSA model with correlated defects with Uniform and Power-law spatial distributions through the introduction of a novel method that is more general than the traditional approach based on master equations dynamics. This method allowed us to elucidate the scaling behavior of fluctuations of coverage fraction from a simple perspective, relying on the CLT. This approach adds new insights into phenomenon of universality breaking raised in \cite{Kundu2021}. It can be concluded that any stochastic process of defect placement that adheres to the assumptions of the CLT will exhibit conventional fluctuations with a scale exponent $\nu=2$. Conversely, if the process does not satisfy these assumptions, the fluctuations will be of the Lévy type, that is, $\nu \neq 2$ and depending on the microscopic details of the model.

\section*{Acknowledgement}
G Palacios thanks a fellowship from Conselho Nacional de Desenvolvimento Científico e Tecnológico (CNPq) Process: 381191/2022-2 and Research grant of Office of Naval Research Global No. N62909-23-1-2014. A. M. S. Mac\^edo acknowledges financial support from Conselho Nacional de Desenvolvimento Científico e Tecnológico - CNPq Grant No. 307626/2022-9. S.K. acknowledges financial support from PNRR Grant CN\_00000013\_CN-HPC, M4C2I1.4, spoke 7, funded by NextGenerationEU. M A F Gomes acknowledges the financial support from the Brazilian Agency CAPES PROEX 23038.003069/2022-87, no. 0041/2022.

\bibliography{rsa}

\begin{thebibliography}{29}%
\makeatletter
\providecommand \@ifxundefined [1]{%
 \@ifx{#1\undefined}
}%
\providecommand \@ifnum [1]{%
 \ifnum #1\expandafter \@firstoftwo
 \else \expandafter \@secondoftwo
 \fi
}%
\providecommand \@ifx [1]{%
 \ifx #1\expandafter \@firstoftwo
 \else \expandafter \@secondoftwo
 \fi
}%
\providecommand \natexlab [1]{#1}%
\providecommand \enquote  [1]{``#1''}%
\providecommand \bibnamefont  [1]{#1}%
\providecommand \bibfnamefont [1]{#1}%
\providecommand \citenamefont [1]{#1}%
\providecommand \href@noop [0]{\@secondoftwo}%
\providecommand \href [0]{\begingroup \@sanitize@url \@href}%
\providecommand \@href[1]{\@@startlink{#1}\@@href}%
\providecommand \@@href[1]{\endgroup#1\@@endlink}%
\providecommand \@sanitize@url [0]{\catcode `\\12\catcode `\$12\catcode
  `\&12\catcode `\#12\catcode `\^12\catcode `\_12\catcode `\%12\relax}%
\providecommand \@@startlink[1]{}%
\providecommand \@@endlink[0]{}%
\providecommand \url  [0]{\begingroup\@sanitize@url \@url }%
\providecommand \@url [1]{\endgroup\@href {#1}{\urlprefix }}%
\providecommand \urlprefix  [0]{URL }%
\providecommand \Eprint [0]{\href }%
\providecommand \doibase [0]{https://doi.org/}%
\providecommand \selectlanguage [0]{\@gobble}%
\providecommand \bibinfo  [0]{\@secondoftwo}%
\providecommand \bibfield  [0]{\@secondoftwo}%
\providecommand \translation [1]{[#1]}%
\providecommand \BibitemOpen [0]{}%
\providecommand \bibitemStop [0]{}%
\providecommand \bibitemNoStop [0]{.\EOS\space}%
\providecommand \EOS [0]{\spacefactor3000\relax}%
\providecommand \BibitemShut  [1]{\csname bibitem#1\endcsname}%
\let\auto@bib@innerbib\@empty
\bibitem [{\citenamefont {Talbot}\ \emph {et~al.}(2000)\citenamefont {Talbot},
  \citenamefont {Tarjus}, \citenamefont {Van~Tassel},\ and\ \citenamefont
  {Viot}}]{talbot2000car}%
  \BibitemOpen
  \bibfield  {author} {\bibinfo {author} {\bibfnamefont {J.}~\bibnamefont
  {Talbot}}, \bibinfo {author} {\bibfnamefont {G.}~\bibnamefont {Tarjus}},
  \bibinfo {author} {\bibfnamefont {P.}~\bibnamefont {Van~Tassel}},\ and\
  \bibinfo {author} {\bibfnamefont {P.}~\bibnamefont {Viot}},\ }\bibfield
  {title} {\bibinfo {title} {From car parking to protein adsorption: an
  overview of sequential adsorption processes},\ }\href
  {https://doi.org/10.1016/S0927-7757(99)00409-4} {\bibfield  {journal}
  {\bibinfo  {journal} {Colloids and Surfaces A: Physicochemical and
  Engineering Aspects}\ }\textbf {\bibinfo {volume} {165}},\ \bibinfo {pages}
  {287} (\bibinfo {year} {2000})}\BibitemShut {NoStop}%
\bibitem [{\citenamefont {Zhang}\ and\ \citenamefont
  {Sun}(2010)}]{zhang2010molecular}%
  \BibitemOpen
  \bibfield  {author} {\bibinfo {author} {\bibfnamefont {L.}~\bibnamefont
  {Zhang}}\ and\ \bibinfo {author} {\bibfnamefont {Y.}~\bibnamefont {Sun}},\
  }\bibfield  {title} {\bibinfo {title} {Molecular simulation of adsorption and
  its implications to protein chromatography: A review},\ }\href
  {https://doi.org/10.1016/j.bej.2009.12.003} {\bibfield  {journal} {\bibinfo
  {journal} {Biochemical Engineering Journal}\ }\textbf {\bibinfo {volume}
  {48}},\ \bibinfo {pages} {408} (\bibinfo {year} {2010})}\BibitemShut
  {NoStop}%
\bibitem [{\citenamefont {Adamczyk}(2012)}]{adamczyk2012modeling}%
  \BibitemOpen
  \bibfield  {author} {\bibinfo {author} {\bibfnamefont {Z.}~\bibnamefont
  {Adamczyk}},\ }\bibfield  {title} {\bibinfo {title} {Modeling adsorption of
  colloids and proteins},\ }\href {https://doi.org/10.1016/j.cocis.2011.12.002}
  {\bibfield  {journal} {\bibinfo  {journal} {Current opinion in colloid \&
  interface science}\ }\textbf {\bibinfo {volume} {17}},\ \bibinfo {pages}
  {173} (\bibinfo {year} {2012})}\BibitemShut {NoStop}%
\bibitem [{\citenamefont {Feder}(1980)}]{Feder1980}%
  \BibitemOpen
  \bibfield  {author} {\bibinfo {author} {\bibfnamefont {J.}~\bibnamefont
  {Feder}},\ }\bibfield  {title} {\bibinfo {title} {Random sequential
  adsorption},\ }\href
  {https://doi.org/https://doi.org/10.1016/0022-5193(80)90358-6} {\bibfield
  {journal} {\bibinfo  {journal} {Journal of Theoretical Biology}\ }\textbf
  {\bibinfo {volume} {87}},\ \bibinfo {pages} {237} (\bibinfo {year}
  {1980})}\BibitemShut {NoStop}%
\bibitem [{\citenamefont {Hinrichsen}\ \emph {et~al.}(1986)\citenamefont
  {Hinrichsen}, \citenamefont {Feder},\ and\ \citenamefont
  {J{\o}ssang}}]{hinrichsen1986geometry}%
  \BibitemOpen
  \bibfield  {author} {\bibinfo {author} {\bibfnamefont {E.~L.}\ \bibnamefont
  {Hinrichsen}}, \bibinfo {author} {\bibfnamefont {J.}~\bibnamefont {Feder}},\
  and\ \bibinfo {author} {\bibfnamefont {T.}~\bibnamefont {J{\o}ssang}},\
  }\bibfield  {title} {\bibinfo {title} {Geometry of random sequential
  adsorption},\ }\href {https://doi.org/10.1007/BF01011908} {\bibfield
  {journal} {\bibinfo  {journal} {Journal of Statistical Physics}\ }\textbf
  {\bibinfo {volume} {44}},\ \bibinfo {pages} {793} (\bibinfo {year}
  {1986})}\BibitemShut {NoStop}%
\bibitem [{\citenamefont {Albano}\ and\ \citenamefont
  {Pereyra}(1993)}]{albano1993adsorption}%
  \BibitemOpen
  \bibfield  {author} {\bibinfo {author} {\bibfnamefont {E.~V.}\ \bibnamefont
  {Albano}}\ and\ \bibinfo {author} {\bibfnamefont {V.~D.}\ \bibnamefont
  {Pereyra}},\ }\bibfield  {title} {\bibinfo {title} {Adsorption kinetics of
  ‘‘hot’’dimers},\ }\href {https://doi.org/10.1063/1.464437} {\bibfield
   {journal} {\bibinfo  {journal} {The Journal of Chemical Physics}\ }\textbf
  {\bibinfo {volume} {98}},\ \bibinfo {pages} {10044} (\bibinfo {year}
  {1993})}\BibitemShut {NoStop}%
\bibitem [{\citenamefont {Budinski-Petkovi{\'c}}\ \emph
  {et~al.}(2017)\citenamefont {Budinski-Petkovi{\'c}}, \citenamefont
  {Lon{\v{c}}arevi{\'c}}, \citenamefont {Dujak}, \citenamefont {Kara{\v{c}}},
  \citenamefont {{\v{S}}{\'c}epanovi{\'c}}, \citenamefont {Jak{\v{s}}i{\'c}},\
  and\ \citenamefont {Vrhovac}}]{budinski2017particle}%
  \BibitemOpen
  \bibfield  {author} {\bibinfo {author} {\bibfnamefont {L.}~\bibnamefont
  {Budinski-Petkovi{\'c}}}, \bibinfo {author} {\bibfnamefont {I.}~\bibnamefont
  {Lon{\v{c}}arevi{\'c}}}, \bibinfo {author} {\bibfnamefont {D.}~\bibnamefont
  {Dujak}}, \bibinfo {author} {\bibfnamefont {A.}~\bibnamefont {Kara{\v{c}}}},
  \bibinfo {author} {\bibfnamefont {J.}~\bibnamefont
  {{\v{S}}{\'c}epanovi{\'c}}}, \bibinfo {author} {\bibfnamefont
  {Z.}~\bibnamefont {Jak{\v{s}}i{\'c}}},\ and\ \bibinfo {author} {\bibfnamefont
  {S.}~\bibnamefont {Vrhovac}},\ }\bibfield  {title} {\bibinfo {title}
  {Particle morphology effects in random sequential adsorption},\ }\href
  {https://doi.org/10.1103/PhysRevE.95.022114} {\bibfield  {journal} {\bibinfo
  {journal} {Physical Review E}\ }\textbf {\bibinfo {volume} {95}},\ \bibinfo
  {pages} {022114} (\bibinfo {year} {2017})}\BibitemShut {NoStop}%
\bibitem [{\citenamefont {Cie{\'s}la}\ \emph {et~al.}(2015)\citenamefont
  {Cie{\'s}la}, \citenamefont {Paja},\ and\ \citenamefont
  {Ziff}}]{ciesla2015shapes}%
  \BibitemOpen
  \bibfield  {author} {\bibinfo {author} {\bibfnamefont {M.}~\bibnamefont
  {Cie{\'s}la}}, \bibinfo {author} {\bibfnamefont {G.}~\bibnamefont {Paja}},\
  and\ \bibinfo {author} {\bibfnamefont {R.~M.}\ \bibnamefont {Ziff}},\
  }\bibfield  {title} {\bibinfo {title} {Shapes for maximal coverage for
  two-dimensional random sequential adsorption},\ }\href@noop {} {\bibfield
  {journal} {\bibinfo  {journal} {Physical Chemistry Chemical Physics}\
  }\textbf {\bibinfo {volume} {17}},\ \bibinfo {pages} {24376} (\bibinfo {year}
  {2015})}\BibitemShut {NoStop}%
\bibitem [{\citenamefont {Kasperek}\ \emph {et~al.}(2018)\citenamefont
  {Kasperek}, \citenamefont {Kubala},\ and\ \citenamefont
  {Cie{\'s}la}}]{kasperek2018random}%
  \BibitemOpen
  \bibfield  {author} {\bibinfo {author} {\bibfnamefont {W.}~\bibnamefont
  {Kasperek}}, \bibinfo {author} {\bibfnamefont {P.}~\bibnamefont {Kubala}},\
  and\ \bibinfo {author} {\bibfnamefont {M.}~\bibnamefont {Cie{\'s}la}},\
  }\bibfield  {title} {\bibinfo {title} {Random sequential adsorption of
  unoriented rectangles at saturation},\ }\href
  {https://doi.org/10.1103/PhysRevE.98.063310} {\bibfield  {journal} {\bibinfo
  {journal} {Physical Review E}\ }\textbf {\bibinfo {volume} {98}},\ \bibinfo
  {pages} {063310} (\bibinfo {year} {2018})}\BibitemShut {NoStop}%
\bibitem [{\citenamefont {Garc{\'\i}a}\ \emph {et~al.}(2015)\citenamefont
  {Garc{\'\i}a}, \citenamefont {Sanchez-Varretti}, \citenamefont {Centres},\
  and\ \citenamefont {Ramirez-Pastor}}]{garcia2015random}%
  \BibitemOpen
  \bibfield  {author} {\bibinfo {author} {\bibfnamefont {G.~D.}\ \bibnamefont
  {Garc{\'\i}a}}, \bibinfo {author} {\bibfnamefont {F.~O.}\ \bibnamefont
  {Sanchez-Varretti}}, \bibinfo {author} {\bibfnamefont {P.~M.}\ \bibnamefont
  {Centres}},\ and\ \bibinfo {author} {\bibfnamefont {A.~J.}\ \bibnamefont
  {Ramirez-Pastor}},\ }\bibfield  {title} {\bibinfo {title} {Random sequential
  adsorption of straight rigid rods on a simple cubic lattice},\ }\href
  {https://doi.org/10.1016/j.physa.2015.05.073} {\bibfield  {journal} {\bibinfo
   {journal} {Physica A: Statistical Mechanics and its Applications}\ }\textbf
  {\bibinfo {volume} {436}},\ \bibinfo {pages} {558} (\bibinfo {year}
  {2015})}\BibitemShut {NoStop}%
\bibitem [{\citenamefont {Ramirez}\ \emph {et~al.}(2023)\citenamefont
  {Ramirez}, \citenamefont {Pasinetti},\ and\ \citenamefont
  {Ramirez-Pastor}}]{ramirez2023random}%
  \BibitemOpen
  \bibfield  {author} {\bibinfo {author} {\bibfnamefont {L.~S.}\ \bibnamefont
  {Ramirez}}, \bibinfo {author} {\bibfnamefont {P.~M.}\ \bibnamefont
  {Pasinetti}},\ and\ \bibinfo {author} {\bibfnamefont {A.~J.}\ \bibnamefont
  {Ramirez-Pastor}},\ }\bibfield  {title} {\bibinfo {title} {Random sequential
  adsorption of self-avoiding chains on two-dimensional lattices},\ }\href
  {https://doi.org/10.1103/PhysRevE.107.064106} {\bibfield  {journal} {\bibinfo
   {journal} {Physical Review E}\ }\textbf {\bibinfo {volume} {107}},\ \bibinfo
  {pages} {064106} (\bibinfo {year} {2023})}\BibitemShut {NoStop}%
\bibitem [{\citenamefont {Krapivsky}(2023)}]{Krapivsky_2023}%
  \BibitemOpen
  \bibfield  {author} {\bibinfo {author} {\bibfnamefont {P.~L.}\ \bibnamefont
  {Krapivsky}},\ }\bibfield  {title} {\bibinfo {title} {Random sequential
  covering},\ }\href {https://doi.org/10.1088/1742-5468/acbc20} {\bibfield
  {journal} {\bibinfo  {journal} {Journal of Statistical Mechanics: Theory and
  Experiment}\ }\textbf {\bibinfo {volume} {2023}},\ \bibinfo {pages} {033202}
  (\bibinfo {year} {2023})}\BibitemShut {NoStop}%
\bibitem [{\citenamefont {Pasinetti}\ \emph {et~al.}(2019)\citenamefont
  {Pasinetti}, \citenamefont {Ram{\'\i}rez}, \citenamefont {Centres},
  \citenamefont {Ramirez-Pastor},\ and\ \citenamefont
  {Cwilich}}]{pasinetti2019random}%
  \BibitemOpen
  \bibfield  {author} {\bibinfo {author} {\bibfnamefont {P.~M.}\ \bibnamefont
  {Pasinetti}}, \bibinfo {author} {\bibfnamefont {L.~S.}\ \bibnamefont
  {Ram{\'\i}rez}}, \bibinfo {author} {\bibfnamefont {P.~M.}\ \bibnamefont
  {Centres}}, \bibinfo {author} {\bibfnamefont {A.~J.}\ \bibnamefont
  {Ramirez-Pastor}},\ and\ \bibinfo {author} {\bibfnamefont {G.~A.}\
  \bibnamefont {Cwilich}},\ }\bibfield  {title} {\bibinfo {title} {Random
  sequential adsorption on euclidean, fractal, and random lattices},\ }\href
  {https://journals.aps.org/pre/abstract/10.1103/PhysRevE.100.052114}
  {\bibfield  {journal} {\bibinfo  {journal} {Physical Review E}\ }\textbf
  {\bibinfo {volume} {100}},\ \bibinfo {pages} {052114} (\bibinfo {year}
  {2019})}\BibitemShut {NoStop}%
\bibitem [{\citenamefont {Ram{\'\i}rez}\ \emph {et~al.}(2019)\citenamefont
  {Ram{\'\i}rez}, \citenamefont {Centres},\ and\ \citenamefont
  {Ramirez-Pastor}}]{ramirez2019inverse}%
  \BibitemOpen
  \bibfield  {author} {\bibinfo {author} {\bibfnamefont {L.~S.}\ \bibnamefont
  {Ram{\'\i}rez}}, \bibinfo {author} {\bibfnamefont {P.~M.}\ \bibnamefont
  {Centres}},\ and\ \bibinfo {author} {\bibfnamefont {A.~J.}\ \bibnamefont
  {Ramirez-Pastor}},\ }\bibfield  {title} {\bibinfo {title} {Inverse
  percolation by removing straight rigid rods from square lattices in the
  presence of impurities},\ }\href {https://doi.org/10.1088/1742-5468/ab054d}
  {\bibfield  {journal} {\bibinfo  {journal} {Journal of Statistical Mechanics:
  Theory and Experiment}\ }\textbf {\bibinfo {volume} {2019}},\ \bibinfo
  {pages} {033207} (\bibinfo {year} {2019})}\BibitemShut {NoStop}%
\bibitem [{\citenamefont {Wang}\ \emph {et~al.}(1993)\citenamefont {Wang},
  \citenamefont {Nielaba},\ and\ \citenamefont {Privman}}]{wang1993locally}%
  \BibitemOpen
  \bibfield  {author} {\bibinfo {author} {\bibfnamefont {J.-S.}\ \bibnamefont
  {Wang}}, \bibinfo {author} {\bibfnamefont {P.}~\bibnamefont {Nielaba}},\ and\
  \bibinfo {author} {\bibfnamefont {V.}~\bibnamefont {Privman}},\ }\bibfield
  {title} {\bibinfo {title} {Locally frozen defects in random sequential
  adsorption with diffusional relaxation},\ }\href
  {https://doi.org/10.1016/0378-4371(93)90066-D} {\bibfield  {journal}
  {\bibinfo  {journal} {Physica A: Statistical Mechanics and its Applications}\
  }\textbf {\bibinfo {volume} {199}},\ \bibinfo {pages} {527} (\bibinfo {year}
  {1993})}\BibitemShut {NoStop}%
\bibitem [{\citenamefont {Ben-Naim}\ and\ \citenamefont
  {Krapivsky}(1994)}]{ben1994irreversible}%
  \BibitemOpen
  \bibfield  {author} {\bibinfo {author} {\bibfnamefont {E.}~\bibnamefont
  {Ben-Naim}}\ and\ \bibinfo {author} {\bibfnamefont {P.}~\bibnamefont
  {Krapivsky}},\ }\bibfield  {title} {\bibinfo {title} {On irreversible
  deposition on disordered substrates},\ }\href
  {https://doi.org/10.1088/0305-4470/27/10/031} {\bibfield  {journal} {\bibinfo
   {journal} {Journal of Physics A: Mathematical and General}\ }\textbf
  {\bibinfo {volume} {27}},\ \bibinfo {pages} {3575} (\bibinfo {year}
  {1994})}\BibitemShut {NoStop}%
\bibitem [{\citenamefont {Tarasevich}\ \emph {et~al.}(2015)\citenamefont
  {Tarasevich}, \citenamefont {Laptev}, \citenamefont {Vygornitskii},\ and\
  \citenamefont {Lebovka}}]{tarasevich2015impact}%
  \BibitemOpen
  \bibfield  {author} {\bibinfo {author} {\bibfnamefont {Y.~Y.}\ \bibnamefont
  {Tarasevich}}, \bibinfo {author} {\bibfnamefont {V.~V.}\ \bibnamefont
  {Laptev}}, \bibinfo {author} {\bibfnamefont {N.~V.}\ \bibnamefont
  {Vygornitskii}},\ and\ \bibinfo {author} {\bibfnamefont {N.~I.}\ \bibnamefont
  {Lebovka}},\ }\bibfield  {title} {\bibinfo {title} {Impact of defects on
  percolation in random sequential adsorption of linear k-mers on square
  lattices},\ }\href {https://doi.org/10.1103/PhysRevE.91.012109} {\bibfield
  {journal} {\bibinfo  {journal} {Physical Review E}\ }\textbf {\bibinfo
  {volume} {91}},\ \bibinfo {pages} {012109} (\bibinfo {year}
  {2015})}\BibitemShut {NoStop}%
\bibitem [{\citenamefont {Centres}\ and\ \citenamefont
  {Ramirez-Pastor}(2015)}]{centres2015percolation}%
  \BibitemOpen
  \bibfield  {author} {\bibinfo {author} {\bibfnamefont {P.~M.}\ \bibnamefont
  {Centres}}\ and\ \bibinfo {author} {\bibfnamefont {A.~J.}\ \bibnamefont
  {Ramirez-Pastor}},\ }\bibfield  {title} {\bibinfo {title} {Percolation and
  jamming in random sequential adsorption of linear k-mers on square lattices
  with the presence of impurities},\ }\href
  {https://doi.org/10.1088/1742-5468/2015/10/P10011} {\bibfield  {journal}
  {\bibinfo  {journal} {Journal of Statistical Mechanics: Theory and
  Experiment}\ }\textbf {\bibinfo {volume} {2015}},\ \bibinfo {pages} {P10011}
  (\bibinfo {year} {2015})}\BibitemShut {NoStop}%
\bibitem [{\citenamefont {Budinski-Petkovi{\'c}}\ \emph
  {et~al.}(2016)\citenamefont {Budinski-Petkovi{\'c}}, \citenamefont
  {Lon{\v{c}}arevi{\'c}}, \citenamefont {Jak{\v{s}}i{\'c}},\ and\ \citenamefont
  {Vrhovac}}]{budinski2016jamming}%
  \BibitemOpen
  \bibfield  {author} {\bibinfo {author} {\bibfnamefont {L.}~\bibnamefont
  {Budinski-Petkovi{\'c}}}, \bibinfo {author} {\bibfnamefont {I.}~\bibnamefont
  {Lon{\v{c}}arevi{\'c}}}, \bibinfo {author} {\bibfnamefont {Z.}~\bibnamefont
  {Jak{\v{s}}i{\'c}}},\ and\ \bibinfo {author} {\bibfnamefont {S.}~\bibnamefont
  {Vrhovac}},\ }\bibfield  {title} {\bibinfo {title} {Jamming and percolation
  in random sequential adsorption of extended objects on a triangular lattice
  with quenched impurities},\ }\href
  {https://doi.org/10.1088/1742-5468/2016/05/053101} {\bibfield  {journal}
  {\bibinfo  {journal} {Journal of Statistical Mechanics: Theory and
  Experiment}\ }\textbf {\bibinfo {volume} {2016}},\ \bibinfo {pages} {053101}
  (\bibinfo {year} {2016})}\BibitemShut {NoStop}%
\bibitem [{\citenamefont {Tarasevich}\ \emph {et~al.}(2017)\citenamefont
  {Tarasevich}, \citenamefont {Laptev}, \citenamefont {Goltseva},\ and\
  \citenamefont {Lebovka}}]{tarasevich2017influence}%
  \BibitemOpen
  \bibfield  {author} {\bibinfo {author} {\bibfnamefont {Y.~Y.}\ \bibnamefont
  {Tarasevich}}, \bibinfo {author} {\bibfnamefont {V.~V.}\ \bibnamefont
  {Laptev}}, \bibinfo {author} {\bibfnamefont {V.~A.}\ \bibnamefont
  {Goltseva}},\ and\ \bibinfo {author} {\bibfnamefont {N.~I.}\ \bibnamefont
  {Lebovka}},\ }\bibfield  {title} {\bibinfo {title} {Influence of defects on
  the effective electrical conductivity of a monolayer produced by random
  sequential adsorption of linear k-mers onto a square lattice},\ }\href
  {https://doi.org/10.1016/j.physa.2017.02.084} {\bibfield  {journal} {\bibinfo
   {journal} {Physica A: Statistical Mechanics and its Applications}\ }\textbf
  {\bibinfo {volume} {477}},\ \bibinfo {pages} {195} (\bibinfo {year}
  {2017})}\BibitemShut {NoStop}%
\bibitem [{\citenamefont {Kundu}\ and\ \citenamefont
  {Mandal}(2021)}]{Kundu2021}%
  \BibitemOpen
  \bibfield  {author} {\bibinfo {author} {\bibfnamefont {S.}~\bibnamefont
  {Kundu}}\ and\ \bibinfo {author} {\bibfnamefont {D.}~\bibnamefont {Mandal}},\
  }\bibfield  {title} {\bibinfo {title} {Breaking universality in random
  sequential adsorption on a square lattice with long-range correlated
  defects},\ }\href {https://doi.org/10.1103/PhysRevE.103.042134} {\bibfield
  {journal} {\bibinfo  {journal} {Phys. Rev. E}\ }\textbf {\bibinfo {volume}
  {103}},\ \bibinfo {pages} {042134} (\bibinfo {year} {2021})}\BibitemShut
  {NoStop}%
\bibitem [{\citenamefont {Palacios}\ and\ \citenamefont
  {Gomes}(2020)}]{palacios2020random}%
  \BibitemOpen
  \bibfield  {author} {\bibinfo {author} {\bibfnamefont {G.}~\bibnamefont
  {Palacios}}\ and\ \bibinfo {author} {\bibfnamefont {M.~A.~F.}\ \bibnamefont
  {Gomes}},\ }\bibfield  {title} {\bibinfo {title} {Random sequential
  adsorption on non-simply connected surfaces},\ }\href
  {https://doi.org/10.1088/1751-8121/ab9fb9} {\bibfield  {journal} {\bibinfo
  {journal} {Journal of Physics A: Mathematical and Theoretical}\ }\textbf
  {\bibinfo {volume} {53}},\ \bibinfo {pages} {375003} (\bibinfo {year}
  {2020})}\BibitemShut {NoStop}%
\bibitem [{\citenamefont {Mantegna}\ and\ \citenamefont
  {Stanley}(1994)}]{mantegna1994stochastic}%
  \BibitemOpen
  \bibfield  {author} {\bibinfo {author} {\bibfnamefont {R.~N.}\ \bibnamefont
  {Mantegna}}\ and\ \bibinfo {author} {\bibfnamefont {H.~E.}\ \bibnamefont
  {Stanley}},\ }\bibfield  {title} {\bibinfo {title} {Stochastic process with
  ultraslow convergence to a gaussian: the truncated {L}{\'e}vy flight},\
  }\href {https://journals.aps.org/prl/abstract/10.1103/PhysRevLett.73.2946}
  {\bibfield  {journal} {\bibinfo  {journal} {Physical Review Letters}\
  }\textbf {\bibinfo {volume} {73}},\ \bibinfo {pages} {2946} (\bibinfo {year}
  {1994})}\BibitemShut {NoStop}%
\bibitem [{\citenamefont {Perc}(2007)}]{perc2007transition}%
  \BibitemOpen
  \bibfield  {author} {\bibinfo {author} {\bibfnamefont {M.}~\bibnamefont
  {Perc}},\ }\bibfield  {title} {\bibinfo {title} {Transition from gaussian to
  {L}{\'e}vy distributions of stochastic payoff variations in the spatial
  prisoner’s dilemma game},\ }\href
  {https://doi.org/10.1103/PhysRevE.75.022101} {\bibfield  {journal} {\bibinfo
  {journal} {Physical Review E}\ }\textbf {\bibinfo {volume} {75}},\ \bibinfo
  {pages} {022101} (\bibinfo {year} {2007})}\BibitemShut {NoStop}%
\bibitem [{\citenamefont {Palacios}\ \emph {et~al.}(2023)\citenamefont
  {Palacios}, \citenamefont {Siqueira}, \citenamefont {de~Ara{\'u}jo},\ and\
  \citenamefont {Raposo}}]{palacios2023replica}%
  \BibitemOpen
  \bibfield  {author} {\bibinfo {author} {\bibfnamefont {G.}~\bibnamefont
  {Palacios}}, \bibinfo {author} {\bibfnamefont {A.~C.}\ \bibnamefont
  {Siqueira}}, \bibinfo {author} {\bibfnamefont {A.~S.~L.}\ \bibnamefont
  {de~Ara{\'u}jo}, \bibfnamefont {Cid B~Gomes}},\ and\ \bibinfo {author}
  {\bibfnamefont {E.~P.}\ \bibnamefont {Raposo}},\ }\bibfield  {title}
  {\bibinfo {title} {Replica symmetry breaking in random lasers: A monte carlo
  study with mean-field interacting photonic random walkers},\ }\href
  {https://doi.org/10.1103/PhysRevA.107.063510} {\bibfield  {journal} {\bibinfo
   {journal} {Physical Review A}\ }\textbf {\bibinfo {volume} {107}},\ \bibinfo
  {pages} {063510} (\bibinfo {year} {2023})}\BibitemShut {NoStop}%
\bibitem [{\citenamefont {Lebovka}\ \emph {et~al.}(2015)\citenamefont
  {Lebovka}, \citenamefont {Tarasevich}, \citenamefont {Dubinin}, \citenamefont
  {Laptev},\ and\ \citenamefont {Vygornitskii}}]{lebovka2015jamming}%
  \BibitemOpen
  \bibfield  {author} {\bibinfo {author} {\bibfnamefont {N.~I.}\ \bibnamefont
  {Lebovka}}, \bibinfo {author} {\bibfnamefont {Y.~Y.}\ \bibnamefont
  {Tarasevich}}, \bibinfo {author} {\bibfnamefont {D.~O.}\ \bibnamefont
  {Dubinin}}, \bibinfo {author} {\bibfnamefont {V.~V.}\ \bibnamefont
  {Laptev}},\ and\ \bibinfo {author} {\bibfnamefont {N.~V.}\ \bibnamefont
  {Vygornitskii}},\ }\bibfield  {title} {\bibinfo {title} {Jamming and
  percolation in generalized models of random sequential adsorption of linear
  k-mers on a square lattice},\ }\href
  {https://doi.org/10.1103/PhysRevE.92.062116} {\bibfield  {journal} {\bibinfo
  {journal} {Physical Review E}\ }\textbf {\bibinfo {volume} {92}},\ \bibinfo
  {pages} {062116} (\bibinfo {year} {2015})}\BibitemShut {NoStop}%
\bibitem [{\citenamefont {Kundu}\ \emph {et~al.}(2022)\citenamefont {Kundu},
  \citenamefont {Prates},\ and\ \citenamefont {Ara{\'{u}}jo}}]{Kundu2022}%
  \BibitemOpen
  \bibfield  {author} {\bibinfo {author} {\bibfnamefont {S.}~\bibnamefont
  {Kundu}}, \bibinfo {author} {\bibfnamefont {H.~C.}\ \bibnamefont {Prates}},\
  and\ \bibinfo {author} {\bibfnamefont {N.~A.~M.}\ \bibnamefont
  {Ara{\'{u}}jo}},\ }\bibfield  {title} {\bibinfo {title} {Jamming and
  percolation in the random sequential adsorption of a binary mixture on the
  square lattice},\ }\href {https://doi.org/10.1088/1751-8121/ac6241}
  {\bibfield  {journal} {\bibinfo  {journal} {Journal of Physics A:
  Mathematical and Theoretical}\ }\textbf {\bibinfo {volume} {55}},\ \bibinfo
  {pages} {204005} (\bibinfo {year} {2022})}\BibitemShut {NoStop}%
\bibitem [{\citenamefont {Mandelbrot}(1982)}]{mandelbrot1982fractal}%
  \BibitemOpen
  \bibfield  {author} {\bibinfo {author} {\bibfnamefont {B.~B.}\ \bibnamefont
  {Mandelbrot}},\ }\href@noop {} {\emph {\bibinfo {title} {The fractal geometry
  of nature}}}\ (\bibinfo  {publisher} {WH freeman New York},\ \bibinfo {year}
  {1982})\BibitemShut {NoStop}%
\bibitem [{\citenamefont {Feller}(1991)}]{feller1991introduction}%
  \BibitemOpen
  \bibfield  {author} {\bibinfo {author} {\bibfnamefont {W.}~\bibnamefont
  {Feller}},\ }\href@noop {} {\emph {\bibinfo {title} {An introduction to
  probability theory and its applications, Volume 1}}},\ Vol.~\bibinfo {volume}
  {81}\ (\bibinfo  {publisher} {John Wiley \& Sons},\ \bibinfo {year}
  {1991})\BibitemShut {NoStop}%
\end{thebibliography}%

\end{document}